\documentclass{ws-p8-50x6-00}

\def\Journal#1#2#3#4{{#1} {\bf #2}, #3 (#4)}

\def\PRL{\em Phys. Rev. Lett.}

\def\ARAA{\em Ann. Rev. Astron. Astrophys.} 
\def\APJ{\em Ap.J.}
\def\APJL{\em Ap.J. Letters}
\def\APJS{\em Ap.J. Suppl.}
\def\AA{\em Astron. Astrophys.}

\def\NA{\em New Astronomy}
\def\AJ{\em AJ}

\def\NATURE{\em Nature}
\def\SSRV{\em Space Science Rev.}

\def\be{\begin{equation}}
\def\ee{\end{equation}}
\def\bea{\begin{eqnarray}}
\def\eea{\end{eqnarray}}

\begin{document}

\title{D/H MEASUREMENTS}

\author{A. VIDAL-MADJAR}

\address{Institut d'Astrophysique de Paris, C.N.R.S./Paris~VI, 98$^{bis}$
Boulevard Arago, 
F-75014, Paris, FRANCE\\E-mail: alfred@iap.fr}

\maketitle

\abstracts{Primordial evaluations of the deuterium abundance should provide
one of the best tests of Big Bang nucleosynthesis models. Space as well as 
ground based observations seem however to result in different values. 
This asks for more observations in different astrophysical sites in order to 
link present day interstellar medium D/H evaluations to primordial ones.
New investigations, made with FUSE (the 
{\it Far Ultraviolet Spectroscopic 
Explorer} launched in June 1999), are presented 
and in the case of the white dwarf G191-B2B line of sight a low D/H
evaluation of 1.16$\pm0.24\times10^{-5}$ (2$\sigma$) is confirmed. This seems to 
indicate that D/H variations are probably present
in the nearby interstellar medium. The FUSE observations 
should help us reach in a near future a better global 
view of the evolution of that key element. }

\section{Introduction}

During 
primordial Big Bang nucleosynthesis deuterium is 
produced in significant amounts 
and then destroyed in stellar 
interiors. It is thus a key element in cosmology and in galactic chemical 
evolution (see {\it e.g.} Audouze \& Tinsley~\cite{at}; 
Boesgaard \& Steigman~\cite{bs}; Olive {\it et~al.}~\cite{oa}; 
Pagel {\it et~al.}~\cite{pa}; 
Vangioni-Flam \& Cass\'e~\cite{vc4}$^{,~}$\cite{vc5}; Prantzos~\cite{p}; 
Scully {\it et~al.}~\cite{sc}; Cass\'e \& Vangioni-Flam~\cite{cv}).

The {\it Copernicus} space observatory has provided the first 
direct measurement of the D/H ratio in the interstellar medium (ISM)
representative of the present epoch (Rogerson \&
York~\cite{ry})~: 

\centerline{(D/H)$^{Copernicus}_{\rm ISM}\simeq1.4\pm0.2\times10^{-5}$.}

More recently D/H evaluations were made in the direction of quasars (QSOs) in
low metallicity media. They were completed
toward three different QSOs' 
(Burles \& Tytler~\cite{bta}$^{,~}$\cite{btb}; 
O'Meara \& Tytler~\cite{ot}) leading to a 
possible range of $2.4-4.8\times10^{-5}$~for the primordial D/H. 
These 
values correspond to a new estimations of the baryon density of the Universe,
$\Omega_{\rm b}{\rm h}^{2}=0.019\pm0.0009$,
in the frame of the 
standard BBN model 
(Burles {\it et~al.}~\cite{bal}; Nollett \& Burles~\cite{nb}). 
When compared to
the recent $\Omega_{\rm b}{\rm h}^{2}$ evaluation made from the 
Cosmic Microwave Background (CMB)
observations (see {\it e.g.} Jaffe {\it et~al.}~\cite{ja})  
$\Omega_{\rm b}{\rm h}^{2}=0.032\pm0.005$, this seems to lead to 
a possible conflict.
Note that another D/H
measurement made toward a low redshift QSO leading to a
D/H value possibly larger than
$10^{-4}$ (Webb {\it et~al.}~\cite{we}; Tytler {\it et~al.}~\cite{ty}) 
corresponds to 
an even stronger disagreement since it translates into
$\Omega_{\rm b}{\rm h}^{2}\le0.01$. 

It is thus important to
investigate the possibility of varying D/H ratios
in different astrophysical sites (see {\it e.g.} Lemoine 
{\it et~al.}~\cite{la9}). 
If variations are indeed found, 
their cause should be investigated before a reliable 
primordial D/H evaluation can be inferred from a small number of observations.

\section{Interstellar observations}

Several methods have been used to measure the interstellar 
D/H ratio. 
All will not be discussed here and for more details see {\it e.g.}
Ferlet~\cite{f2}.
The more reliable approach is to observe in absorption, against the 
background continuum
of stars, the atomic Lyman series of D and H 
in the far-UV. 

Toward hot stars,  with the {\it Copernicus}
satellite, many important evaluations of D/H were obtained
(see e.g. Rogerson and York~\cite{ry}; York and Rogerson~\cite{yr}; 
Vidal--Madjar {\it et~al.}~\cite{va1977};
Laurent {\it et~al.}~\cite{lvy}; Ferlet {\it et~al.}~\cite{fa1980}; 
York~\cite{y1983}; Allen {\it et~al.}~\cite{aa1992})
leading to the detection of variations 
recently enforced by HST--GHRS (Vidal--Madjar {\it et~al.}~\cite{va})
toward G191--B2B showing a low value and IMAPS observations, one made 
toward 
$\delta$~Ori presenting again a low value (Jenkins {\it et~al.}~\cite{ja})
confirming the previous analysis made by Laurent {\it et~al.}~\cite{lvy} 
from {\it Copernicus} observations and the 
other one 
toward $\gamma^2$~Vel with a high value (Sonneborn {\it et~al.}~\cite{so}).
These observations seem to indicate that in the ISM, within few hundred
parsecs, D/H may vary by more
than a factor $\simeq3$.

From published values, D/H ranges from

\centerline{$\sim5\times10^{-6}$~$<$~(D/H)$_{ISM}$~$<$~$\sim4\times10^{-5}$.}

This method also provided a precise 
D/H evaluation in the local ISM (LISM) in the direction of the cool star
Capella 
(Linsky {\it et~al.}~\cite{la})~:

\centerline{(D/H)$^{\rm GHRS}_{\rm Capella}=1.60\pm0.09^{+0.05}_{-0.10}\times10^{-5}$} 

Additional observations made in the LISM lead Linsky~\cite{l98} (see
references there in) to the conclusion that the D/H
value within the Local Interstellar Cloud (LIC) is 
(compatible with 12 evaluations)~:

\centerline{(D/H)$^{\rm GHRS}_{\rm LIC}=1.50\pm0.10\times10^{-5}$} 

\section{The nearby ISM}

Observations of white dwarfs (WD) in the nearby ISM (NISM)
for precise D/H evaluations were first proposed and
achieved in the direction of G191--B2B by Lemoine {\it et~al.}~\cite{la6} using
the HST--GHRS spectrograph at medium resolution. Follow up 
observations on G191--B2B at higher resolution with the GHRS
Echelle-A grating by Vidal--Madjar {\it et~al.}~\cite{va}
(same instrument configuration used as in the Capella study) 
lead to a precise D/H evaluation in the NISM 
along this line of sight within 
one H{\sc i} region -- the Local Interstellar Cloud
(LIC) also observed toward Capella 
(these stars are separated by $\sim7^{o}$ on the sky)
-- and within a more complex and
ionized H{\sc ii} region presenting a double velocity structure. In these two
main interstellar components the D/H ratio was found to be
different if one assumes that the D/H value within the LIC is
the same as the one found in the direction of Capella, in which case 
D/H has to be lower ($\sim0.9\times10^{-5}$) in the more ionized 
components.

In any case a lower ``average'' D/H ratio is found 
(2$\sigma$ error)~:

\centerline{(D/H)$^{\rm GHRS}_{\rm G191-B2B}=1.12\pm0.16\times10^{-5}$} 

\begin{figure*}
\center
\psfig{figure=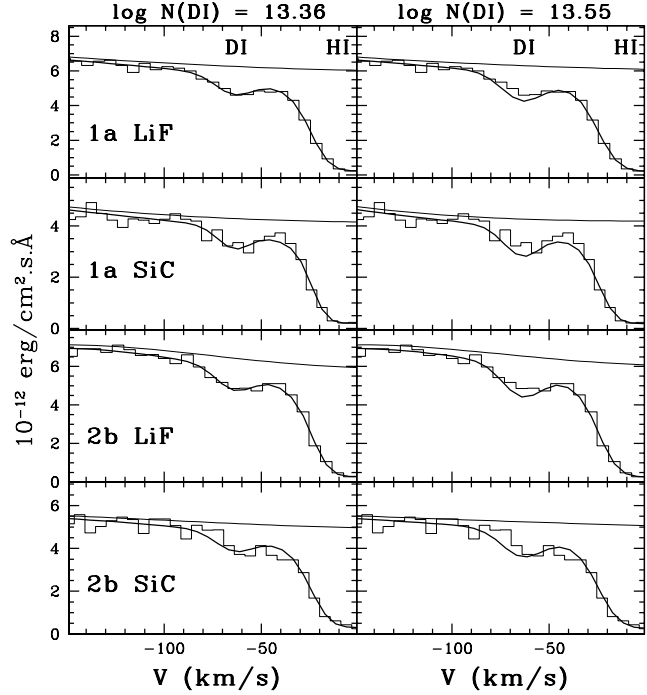,height=3.8in}
\caption{FUSE observations made in the direction of G191--B2B. The
Lyman~$\beta$ line is shown in the four FUSE channels recorded
simultaneously through the high resolution slit (two LiF and two SiC). On the
left panels the fits are shown with the best FUSE evaluation made by
Vidal--Madjar {\it et al.}~$^{35}$ 
and on the
right ones with the D column density evaluated by Sahu {\it et al.}~$^{33}$
with the STIS instrumentation which is more than 6$\sigma$ incompatible
with the FUSE observations.}
\label{fig:gbb}
\end{figure*}

This result has been 
contested by Sahu {\it et~al.}~\cite{sa9} who used new HST--STIS high
resolution Echelle observations. However Vidal--Madjar~\cite{v}
has showed that all data sets (GHRS and STIS)
in fact converge on a same value of the D/H
ratio, which furthermore agrees with that derived by 
Vidal--Madjar {\it et~al.}~\cite{va} and disagrees with that of 
Sahu {\it et~al.}~\cite{sa9}.

Since the disagreement between the two analysis was 
on the D{\sc i} column
density estimation, FUSE observations were 
expected to clarify the situation since they give access to 
weaker deuterium Lyman lines that are
less sensitive to saturation effects than Lyman~$\alpha$.
Three independent data
sets were obtained corresponding to the three different 
FUSE entrance apertures (Vidal--Madjar 
{\it et~al.}~\cite{va1}). 
The fits of
the D Lyman~$\beta$ line in the various 
FUSE channels are shown in Figure~1
and compared with the estimate of Sahu {\it et~al.}~\cite{sa9}.
These new data confirm the measurement of N(D{\sc i}) of Vidal--Madjar 
{\it et~al.}~\cite{va}; the value N(D{\sc i}) derived by
Sahu {\it et~al.}~\cite{sa9} lies 6$\sigma$ away from the new result. These 
6$\sigma$ are quantified in terms of $\Delta\chi^2$, 
including many possible systematics such as stellar continuum 
placement, zero
level, spectral instrument shifts, line spread function profiles, all free
in the fitting process (see e.g. the different stellar 
continuum levels in Figure~1
from left to right panels).

The H{\sc i} column density toward G191--B2B is well determined.
Independent measurements with EUVE (Dupuis {\it et~al.}~\cite{da5}), 
GHRS (medium\cite{la6} and 
high resolution\cite{v}) and STIS (high resolution\cite{sa9})
using several 
methods of evaluation (EUV, Lyman continuum opacity and Lyman~$\alpha$, 
damping wing modelling), 
converge on a value of log~N(H{\sc i})~=~18.34~($\pm0.03$).  
The error on this value includes systematic errors associated with the 
various measurement techniques.

Using the D{\sc i} column
density as measured by FUSE and the H{\sc i} column
density compatible with all published values, one arrives at
(2$\sigma$ error)~: 

\centerline{(D/H)$^{\rm FUSE-HST-EUVE}_{\rm G191-B2B}=1.16\pm0.24\times10^{-5}$} 
 
This value is marginally compatible ($\ge2\sigma$) with the LIC one.

The essential question remains~: 
if D/H variations are confirmed in  
more sightlines, what could be their cause~?

\section{The FUSE observatory}

FUSE starts to produce orders of magnitude more data on the 
distribution of D/H in the ISM. From the planned D/H survey, 
we should be able to 
evaluate the deuterium abundance in a wide variety 
of locations, possibly linked to the past star formation rate as well as
to the supposed infall of less processed gas 
in our Galaxy, and thus better understand Galactic chemical evolution.

The FUSE sensitivity should allow evaluations of the deuterium 
abundance in tens of lines of sights~:
i) in the direction of white dwarfs and 
cool stars in the NISM~;
ii) toward hot sub-dwarfs in the more distant ISM and nearby Galactic halo~;
iii) within the Galactic disk over several kilo-parsecs in the 
direction of O and early B stars~;
iv) in the more distant Galactic halo, within high velocity cloud complexes 
as well as in
intergalactic clouds in the direction of low redshift QSOs, AGNs and blue
compact galaxies.

\begin{figure}
\center
\psfig{figure=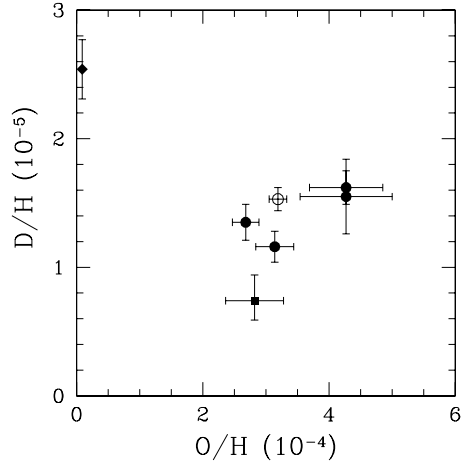,height=2.5in}
\caption{Different D/H evaluations as a function of O/H (1$\sigma$ errors). 
The diamond
is the observation made in the direction of a QSO by O'Meara
and Tytler~$^{13}$ corresponding to a low metallicity cloud;  
the square is the IMAPS observation in the direction of $\delta$Ori by
Jenkins $et~al.~^{28}$ in the ISM; the filled circles correspond to the
FUSE observations (and among them the G191--B2B one)
and the open circle to the average of 12 lines of sight through the LIC
observed by GHRS and STIS 
from Linsky~$^{31}$ in relation to the ISM O/H evaluation
made by Meyer $et~al.~^{43}$ from their survey.}
\label{fig:DsH-OsH}
\end{figure}

The first precise D/H evaluations toward few white dwarfs
were presented in early 2001 
at the AAS meeting
(Moos {\it et~al.}~\cite{ma1}; Friedman {\it et~al.}~\cite{fa1};
H\'ebrard {\it et~al.}~\cite{ha1}; Kruk {\it et~al.}~\cite{ka1};
Linsky {\it et~al.}~\cite{la1}; Sonneborn {\it et~al.}~\cite{sa1}; 
Vidal--Madjar {\it et~al.}~\cite{va1}). 
The deuterium Lyman lines are clearly seen toward these few WDs 
and, as an example, the Lyman~$\beta$ line is shown in the case 
of G191--B2B as previously
discussed (see
Figure~1). 
Several of these D/H evaluations made 
in the ISM with FUSE, HST, IMAPS are shown in Figure~2 
along with one made recently 
in the direction of one
QSO from ground based observations~\cite{ot}, 
as a function of the line of
sight average metallicity as traced by O/H when available. 
It seems that the D/H variation does not anti--correlate 
with O/H.
Thus a simple mechanism as astration, 
able to destroy D and produce O, 
does not seem compatible with the observations. Other
mechanisms should be investigated as the ones listed by {\it e.g.}
Lemoine {\it et~al.}~\cite{la9}.

\section{Conclusion}

In summary the status of
the different -- but discordant -- D/H evaluations taken with no a priori 
bias to select one over another could be the following. 

If the variations of the D/H ratio in the NISM are 
illusory, one could quote an average 
value of 
(D/H)$_{\rm NISM} \simeq 1.3-1.4\times10^{-5}$ barely compatible with all
observations.

More in agreement with the present 
observations, D/H seems to vary in the ISM.
One has thus to understand why. 

Until then, 
any single or small number of values should not be considered to represent
the definitive D/H in a given region.
This is particularly true for 
the ``primordial'' values found in
the direction of QSOs since the physical state of the probed environment is
more poorly known than the Galactic one.  

Our hope is that the FUSE mission will solve these
problems.

\section*{Acknowledgments}
I am very grateful for the entire FUSE operation team for all the
impressive work they are doing to make the FUSE observatory come true.
I also thank the whole FUSE team for many positive interaction and
comments. This work is based on
data obtained by the NASA--CNES--CSA FUSE mission operated by the Johns
Hopkins University under NASA contract NAS5--32985.


\begin{thebibliography}{99}
\bibitem{at}J. Audouze and B.M. Tinsley, \Journal{\ARAA} 
{14}{43}{1976}.
\bibitem{bs}A.M. Boesgaard and G. Steigman, \Journal{\ARAA} 
{23}{319}{1985}.
\bibitem{oa}K. Olive {\it et~al.}, \Journal{\PRL} 
{B236}{454}{1990}.
\bibitem{pa}B.  Pagel {\it et~al.}, \Journal{MNRAS}
{255}{325}{1992}.
\bibitem{vc4}E. Vangioni-Flam and M. Cass\'e, \Journal{\APJ} 
{427}{618}{1994}.
\bibitem{vc5}E. Vangioni-Flam and M. Cass\'e, \Journal{\APJ} 
{441}{471}{1995}.
\bibitem{p}N. Prantzos, \Journal{AA} 
{310}{106}{1996}.
\bibitem{sc}S.T. Scully, {\it et~al.}, \Journal{\APJ} 
{476}{521}{1997}.
\bibitem{cv}M. Cass\'e and E. Vangioni-Flam in {\em Structure and Evolution of 
the Intergalactic Medium from QSO Absorption Line Systems}, eds. P. Petitjean
and S. Charlot (IAP Conference, 331, 1998).
\bibitem{ry}J. Rogerson and D. York, \Journal{\APJL} 
{186}{L95}{1973}.
\bibitem{bta}S. Burles and D. Tytler, \Journal{\APJ} 
{499}{699}{1998a}.
\bibitem{btb}S. Burles and D. Tytler, \Journal{\APJ} 
{507}{732}{1998b}.
\bibitem{ot}J. O'Meara and D. Tytler, in these proceedings
{\em Cosmic Evolution}, eds. M. Lemoine and R. Ferlet, 2001.
\bibitem{bal}S. Burles {\it et~al.}, \Journal{\PRL} 
{82}{4176}{1999}.
\bibitem{nb}K.M. Nollett and S. Burles, \Journal{\PRL} 
{D61}{123505}{2000}.
\bibitem{j}A.H. Jaffe {\it et~al.}, astro-ph/0007333, 2000. 
\bibitem{we}J.K. Webb {\it et~al.}, \Journal{\NATURE} 
{388}{250}{1997}.
\bibitem{ty}D. Tytler {\it et~al.}, \Journal{\AJ} 
{117}{63}{1999}.
\bibitem{la9}M. Lemoine {\it et~al.}, \Journal{\NA} 
{4}{231}{1999}.
\bibitem{f2}R. Ferlet, in IAU$\#$150 {\em Astrochemistry of Cosmic Phenomena},
eds. P.D. Singh, (Kluwer, 85, 1992).        
\bibitem{yr}D. York and J. Rogerson, \Journal{\APJL} 
{203}{378}{1976}.
\bibitem{va1977}A. Vidal--Madjar {\it et al.}, \Journal{\APJL} 
{211}{91}{1977}.
\bibitem{lvy}C. Laurent, A. Vidal--Madjar and D.G. York, \Journal{\APJ} 
{229}{923}{1979}.
\bibitem{fa1980}R. Ferlet {\it et al.}, \Journal{\APJL} 
{242}{576}{1980}.
\bibitem{y1983}D.G. York, \Journal{\APJL} 
{264}{172}{1983}.
\bibitem{aa1992}M.M. Allen, E.B. Jenkins and T.P. Snow, \Journal{\APJS} 
{83}{261}{1992}.
\bibitem{va}A. Vidal--Madjar {\it et~al.}, \Journal{\AA} 
{338}{694}{1998}.
\bibitem{ja}E.B. Jenkins {\it et~al.}, \Journal{\APJ} 
{520}{182}{1999}.
\bibitem{so}G. Sonneborn, {\it et~al.}, \Journal{\APJ} 
{545}{277}{2000}.
\bibitem{la}J. Linsky {\it et~al.}, \Journal{\APJ} 
{451}{335}{1995}.
\bibitem{l98}J. Linsky, \Journal{\SSRV} 
{84}{285}{1998}.
\bibitem{la6}M. Lemoine {\it et~al.}, \Journal{\AA} 
{308}{601}{1996}.
\bibitem{sa9}M.S. Sahu {\it et~al.}, \Journal{\APJL} 
{523}{L159}{1999}.
\bibitem{v}A. Vidal--Madjar, in {\em The Light Elements and
Their Evolution}, eds. L.~da~Silva, M.~Spite
and J.~R.~de~Medeiros (ASP Conference Series, 151, 2000).
\bibitem{va1}A. Vidal--Madjar {\it et~al.}, {\it Ap.J.} in preparation, (2001).
\bibitem{da5}J. Dupuis {\it et~al.}, \Journal{\APJ} 
{455}{574}{1995}.
\bibitem{ma1}H.W. Moos {\it et~al.}, {\it Ap.J.} in preparation, (2001).
\bibitem{fa1}S.D. Friedman {\it et~al.}, {\it Ap.J.} in preparation, (2001).
\bibitem{ha1}G. H\'ebrard {\it et~al.}, {\it Ap.J.} in preparation, (2001).
\bibitem{ka1}J.W. Kruk {\it et~al.}, {\it Ap.J.} in preparation, (2001).
\bibitem{la1}J.L. Linsky {\it et~al.}, {\it Ap.J.} in preparation, (2001).
\bibitem{sa1}G. Sonneborn {\it et~al.}, {\it Ap.J.} in preparation, (2001).
\bibitem{ma8}D.M. Meyer {\it et~al.}, \Journal{\APJ} 
{493}{222}{1998}.

\end{thebibliography}
\end{document}